\documentclass[onecolumn,superscriptaddress,amsmath,amssymb,notitlepage,nofootinbib,showpacs,preprintnumbers]{revtex4-1}

\usepackage{amsfonts}
\usepackage{amsmath}
\usepackage{amssymb}
\usepackage{graphicx}
\usepackage{dcolumn}
\usepackage{bm}
\usepackage{tabularx}
\usepackage{epstopdf}
\usepackage{xcolor}
\usepackage{mathrsfs}
\usepackage{mathtools}
\usepackage{float}
\usepackage{algpseudocode}
\usepackage{verbatim}
\usepackage{wasysym}
\usepackage{footnote}
\usepackage{chngcntr}
\usepackage{natbib}

\begin{document}

\title{Bitcoin Transaction Networks: an overview of recent results}

\author{Nicol\'o Vallarano}
\affiliation{IMT School for Advanced Studies, Piazza S.Francesco 19, 55100 Lucca (Italy)}
\author{Claudio Tessone}
\email{claudio.tessone@business.uzh.ch}
\affiliation{UZH Blockchain Center, University of Z\"urich, R\"amistrasse 71, 8006 Z\"urich (Switzerland)}
\affiliation{URPP Social Networks, University of Z\"urich, Andreasstrasse 15, 8050 Z\"urich (Switzerland)}
\author{Tiziano Squartini}
\affiliation{IMT School for Advanced Studies, Piazza S.Francesco 19, 55100 Lucca (Italy)}
\date{\today}

\begin{abstract}
Cryptocurrencies are distributed systems that allow exchanges of native (and non-) tokens among participants. The availability of the complete historical bookkeeping opens up an unprecedented possibility, i.e. that of understanding the evolution of their network structure while gaining useful insight on the relationships between users' behaviour and cryptocurrency pricing in exchange markets. In this contribution we review some of the most recent results concerning the structural properties of the \emph{Bitcoin Transaction Networks}, a generic name referring to a set of three different constructs: the \emph{Bitcoin Address Network}, the \emph{Bitcoin User Network} and the \emph{Bitcoin Lightning Network}. The picture that emerges is that of system growing over time, which becomes increasingly sparse and whose mesoscopic structural organization is characterised by the presence of an increasingly significant core-periphery structure. Such a peculiar topology is matched by a highly uneven distribution of bitcoins, a result suggesting that Bitcoin is becoming an increasingly centralized system at different levels.
\end{abstract}

\maketitle

\section*{Introduction}

A cryptocurrency is an online payment system for which the storage and the verification of transactions - therefore, the safeguard of the system consistency itself - are \emph{decentralized}, i.e. do not require the presence of a trusted third party. This result can be achieved by securing financial transactions through a clever combination of cryptographic technologies \cite{antonopoulos2017mastering}.

Bitcoin, the first and most popular cryptocurrency, was introduced in 2008 by Satoshi Nakamoto \cite{nakamoto2008bitcoin}: it consists of a decentralized peer-to-peer network to which users connect to exchange the property of the account units of the system, i.e. perform \emph{bitcoins} transactions. Each transaction becomes part of a publicly available ledger, the \emph{blockchain}, after having been validated by the so-called \emph{miners}, i.e. users that verify the validity of issued transactions according to the consensus rules that are part of the Bitcoin protocol \cite{halaburda2016beyond,glaser2017pervasive}. A new block, containing transactions that where known to the miner since the last block, is `mined' - on average - every 10 minutes, thereby adding new transactions to the blockchain. Thus, these transactions are `confirmed', in turn enabling users to spend the bitcoins they received through them\footnote{In practice, the so-called `6 confirmations' rule is followed: once a transaction is included in a block which is followed by at least six additional blocks \cite{decker2013information}, the transaction can be safely considered as confirmed.}. The cryptography protocols Bitcoin rests upon aim at preventing the so-called \emph{double-spending problem}, i.e. the possibility for the same digital token to be spent more than once in absence of a central party that guarantees the validity of the transactions \cite{nakamoto2008bitcoin,antonopoulos2017mastering}: remarkably, the transaction-verification mechanism Bitcoin relies on allows its entire transaction history to be openly accessible, a feature that, in turn, allows researchers to analyse it in different network representation. 

The gain of Bitcoin popularity led its community to face new problems such as the lack of \emph{scalability} of the transaction-verification method, i.e. the (relatively low) maximum number of transactions that can be verified per second - especially if compared with the mainstream competitors such as centralized payment networks, the increasing \emph{concentration of mining power} in mining pools - implying that the verification mechanism in the network becomes increasingly less distributed - and the tendency of users to \emph{hoard}. In order to solve the aforementioned problems, threatening the overall functioning of Bitcoin as a medium of exchange, new instruments were adopted. Proposed in 2015 \cite{poon2016bitcoin}, the \emph{Bitcoin Lightning Network} (BLN) is a `Layer 2' protocol that can operate on top of blockchain-based (Bitcoin-like) cryptocurrencies by creating bilateral channels for \emph{off-chain} payments which are, then, settled concurrently on the blockchain, once the channels get closed. As both the transaction fees and the blockchain confirmation are no longer required, the network is spared from avoidable burden; moreover, the key features of Bitcoin, i.e. its \emph{decentralized architecture}, its \emph{political organisation} and its \emph{wealth distribution} are no longer sacrificed, while the circulation of the native assets is enhanced.

Bitcoin is almost ten years old: however, while a large amount of literature concerning either the purely financial or the purely engineeristic aspect of it indeed exists (e.g. the prediction of the exchange rate of Bitcoin versus the US dollar \cite{hencic2014noncausal}, the statistical properties of the former one \cite{chu2015statistical}, the statistical properties of Bitcoin daily log-returns \cite{chu2017statistical}, the comparison of the Bitcoin volatility with the one of the exchange rates of major global currencies \cite{sapuric2014bitcoin,briere2015virtual}, the identification of factors influencing the Bitcoin price \cite{kristoufek2015price}, its predictability via machine-learning techniques \cite{Yiying2019proceeds}, the interplay between social interactions and the movements of the Bitcoin price \cite{garcia2014digital,elBahrawy2019wikipedia}, the problem of Bitcoin users de-anonymisation \cite{androulaki2013evaluating, harrigan2016unreasonable,meiklejohn2013fistful,ober2013structure,reid2013analysis}), researchers have started to investigate the Bitcoin \emph{structural} properties only recently. In \cite{kondor2014rich}, the authors consider the network of transactions between addresses at the weekly time scale, showing the emergence of power-law distributions and that the number of incoming transactions reflects the wealth of nodes; in \cite{javarone2018from}, the authors consider the network of transactions between users at the macroscale, in order to check for its small-worldness; in \cite{parino2018analysis}, the authors investigate the network of international Bitcoin flows, identifying socio-economic factors that drive its adoption across countries. In general, however, the contributions analyzing Bitcoin from a network perspective provide a quite limited overview of its evolution, either focusing on a single representation of it or on a relatively short period of time; even the ones addressing the problem from a wider perspective \cite{motamed2019quantitative,liang2018evolutionary} often limit themselves to a purely descriptive analysis, without comparing the empirical observations with the outcome of proper models.

With the present work, we aim at summing up the results of three papers both providing a comprehensive overview of the empirical traits characterizing the Bitcoin evolution and framing them within models rooted into statistical physics \cite{bovet2019evolving,vallarano2020exploring,lin2020lightning}. In \cite{bovet2019evolving}, the authors analyse the local properties of two different Bitcoin representations, i.e. the \emph{Bitcoin Address Network} (BAN) and the \emph{Bitcoin User Network} (BUN) and inspect the presence of correlations between (exogenous) price movements and (endogenous) changes in the topological structure of the aforementioned networks. In \cite{vallarano2020exploring}, the mesocale structure of the BUN is under scrutiny: particular attention is devoted to the identification of the best network model able to describe it; besides, the same exercise as above is carried out, i.e. the comparison between the evolution of purely structural properties and the appearance of price bubbles in a cyclical fashion. Lastly, in \cite{lin2020lightning}, the authors inspect the evolution of the BLN topology, pointing out that it is becoming an increasingly centralized system and that the `capital' is becoming increasingly unevenly distributed.

\section*{Data}

As previously said, Bitcoin relies on a decentralized public ledger, the blockchain, that records all transactions among Bitcoin users. A transaction is a set of input and output addresses: the output addresses that are `unspent', i.e. not yet recorded on the ledger as input addresses, can be claimed, and therefore spent, only by the owner of the corresponding cryptographic key. This is the reason why one speaks of \emph{pseudonimity}: an observer of the blockchain can see all unspent \emph{addresses} but cannot link them to the actual owners.

\subsection*{The Bitcoin Address Network (BAN)}

The BAN is the simplest network that can be constructed from the blockchain records: from a technical point of view, it is a directed, weighted graph whose nodes represent addresses; the direction and the weight of links are provided by the input-output relationships defining the transactions recorded on the blockchain. The BAN has been considered across a period of 9 years, i.e. from 9th Janury 2009 to 18th December 2017 at the end of which the data set consists of 304\:111\:529 addresses, exchanging a total number of transactions amounting at 283\:028\:575. In terms of traded volume, the transactions between addresses amount at 4\:432\:597\:496 bitcoins.

\subsection*{The Bitcoin User Network (BUN)}

Since the same owner may control several addresses \cite{meiklejohn2013fistful}, one can define a network of `users' whose nodes are \emph{clusters of addresses}. These clusters are derived by implementing different \textit{heuristics}, provided by the state-of-the-art literature \cite{androulaki2013evaluating,harrigan2016unreasonable,tasca2018evolution,ron2013quantitative}. The `user networks' we obtain should not be considered as a perfect representation of the actual networks of users but, rather, an attempt to group addresses while minimising the presence of false positives. Two heuristics have been employed, here: the multi-input one (based on the assumption that the addresses appearing as input of the same transaction are controlled by the same user) and the change address one (based on the assumption that a new address appearing as output of a transaction and with the lowest amount of transferred money must belong to the input user). The BUN has been considered across the same period of the BAN (i.e. of 9 years, from 9th Janury 2009 to 18th December 2017) at the end of which the data set consists of 16\:749\:939 users, exchanging a total number of transactions amounting at 224\:620\:265. In terms of traded volume, the transactions between users amount at 3\:114\:359\:679 bitcoins.

\subsection*{The Bitcoin Lightning Network (BLN)}

The BLN is constructed in a fashion that is similar to way the BAN is defined: it is a directed, weighted graph whose nodes are the addresses exchanging bitcoins on the `Layer 2'. Three different representations of the BLN have been studied so far, i.e. the daily one, the weekly one and the daily-block one: while a daily/weekly snapshot includes all channels that were found to be active during that day/week, a daily-block snapshot consists of all channels that were found to be active at the time the first block of the day was released (hence, the transactions considered for the daily-block representation are a subset of the ones constituting the daily representation). The BLN was considered across a period of 18 months, i.e. from 14th January 2018 to 13th July 2019, at the end of which the network consists of 8\:216 users, 122\:517 active channels and 2\:732,5 transacted bitcoins.

\subsection*{Notation}

Although the information about the magnitude of transactions is available, the BAN and the BUN have been analysed as binary, directed networks; as such, they are completely specified by their binary, asymmetric adjacency matrices $\mathbf{A}_\text{BAN}^{(t)}$ and $\mathbf{A}_\text{BUN}^{(t)}$, at time $t$. The generic entry $a_{ij}^{(t)}$ is equal to 1 if at least one transaction between address (user) $i$ and address (user) $j$ takes place, i.e. bitcoins are transferred from address (user) $i$ to address (user) $j$, during the time snapshot $t$ and 0 otherwise. The BLN, on the other hand, is a weighted, undirected network, represented by a symmetric matrix $\mathbf{W}_\text{BLN}^{(t)}$ whose generic entry $w_{ij}^{(t)}=w_{ji}^{(t)}$ indicates the total amount of money exchanged between $i$ and $j$, across all channels, at time $t$; here, we will mainly focus on its binary projection $\mathbf{B}_\text{BLN}^{(t)}$, whose generic entry reads $b_{ij}^{(t)}=b_{ji}^{(t)}=1$ if $w_{ij}^{(t)}=w_{ji}^{(t)}>0$ and $b_{ij}^{(t)}=b_{ji}^{(t)}=0$ otherwise.

\section*{Results}

\subsection*{The Bitcoin Address Network and the Bitcoin User Network}

Let us start by reviewing the results concerning the BAN and the BUN at the weekly time scale. Similar results are observed for the BAN and the BUN at the daily time scale \cite{bovet2019evolving}.\\

\begin{figure*}[t!]
\centering
\includegraphics[width=\textwidth]{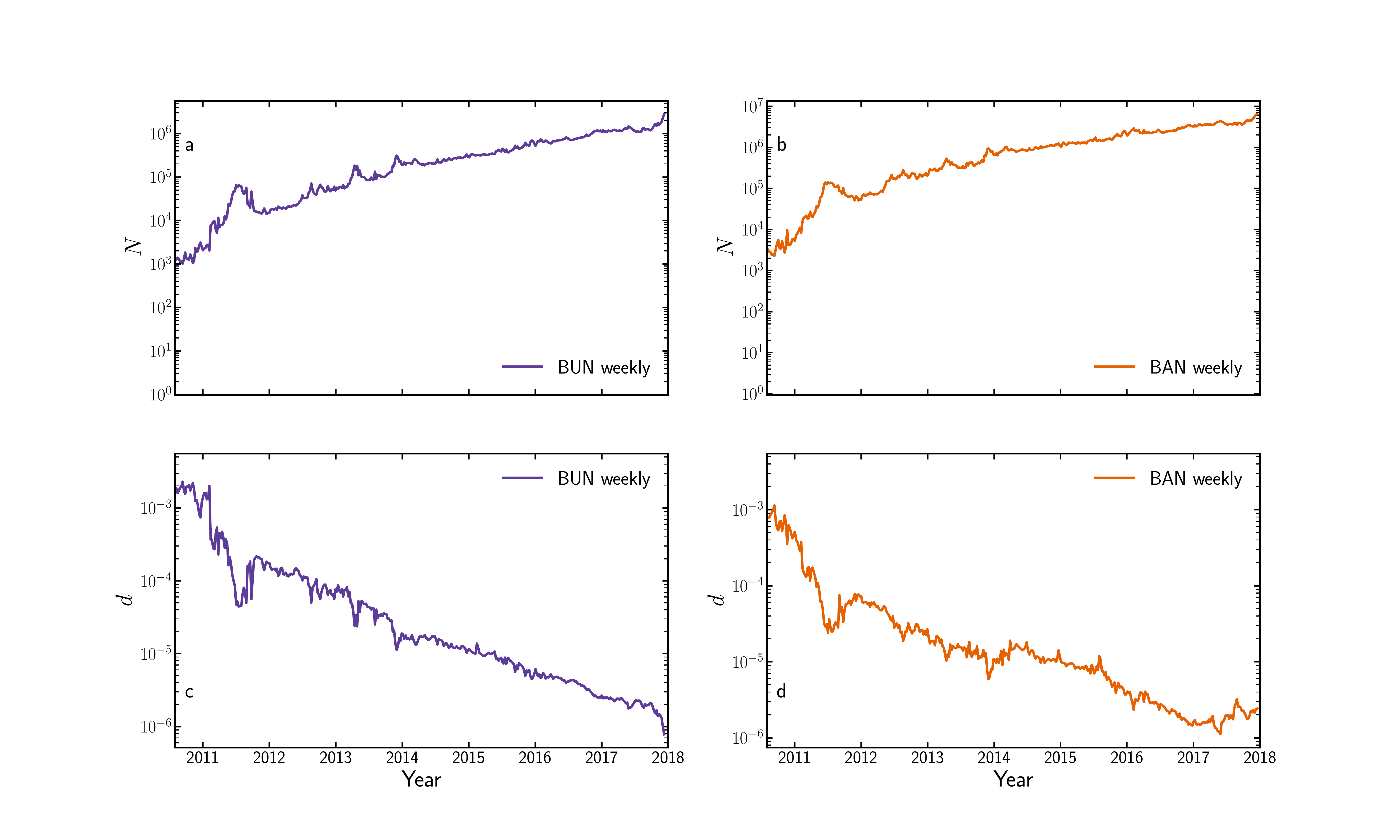}
\caption{Evolution of basic statistics, i.e. the number of nodes (left panels) and the link density (right panels) for two Bitcoin network representations, i.e. the BAN (bottom panels) and the BUN (top panels) at the weekly time scale, from July 2010 to 18th December 2017 (i.e. for networks with at least 200 nodes). Although the network size increases, it becomes sparser (irrespectively from the considered representation). Similar results are observed for the BAN and the BUN at the daily time scale. Adapted from \cite{bovet2019evolving}.}
\label{fig:1}
\end{figure*}

\paragraph*{Basic statistics.} Let us start by commenting one the evolution of some basic statistics characterising the BAN and the BUN that, as noticed elsewhere \cite{ober2013structure}, have started to evolve in a more stationary fashion since middle 2011. As fig. \ref{fig:1} shows, both the number of nodes $N$ and the number of links $L=\sum_i\sum_{j(\neq i)}a_{ij}$ increase steadily in time, irrespectively from the considered representation; the link density $d=\frac{L}{N(N-1)}$, however, decreases, meaning that the system becomes sparser. The dependence of $d$ from $N$ can be better specified, from a mathematical point of view, upon noticing that the average degree $\overline{k^{in}}=\overline{k^{out}}=\frac{\sum_i\sum_{j(\neq i)}a_{ij}}{L}=\frac{L}{N}$ is either constant (for what concerns the BUN) or limited (for what concerns the BAN) over time \cite{bovet2019evolving}; hence, it follows that $L\propto N$ and $d\sim N^{-1}$.\\

\begin{figure*}[t!]
\centering
\includegraphics[width=0.9\textwidth]{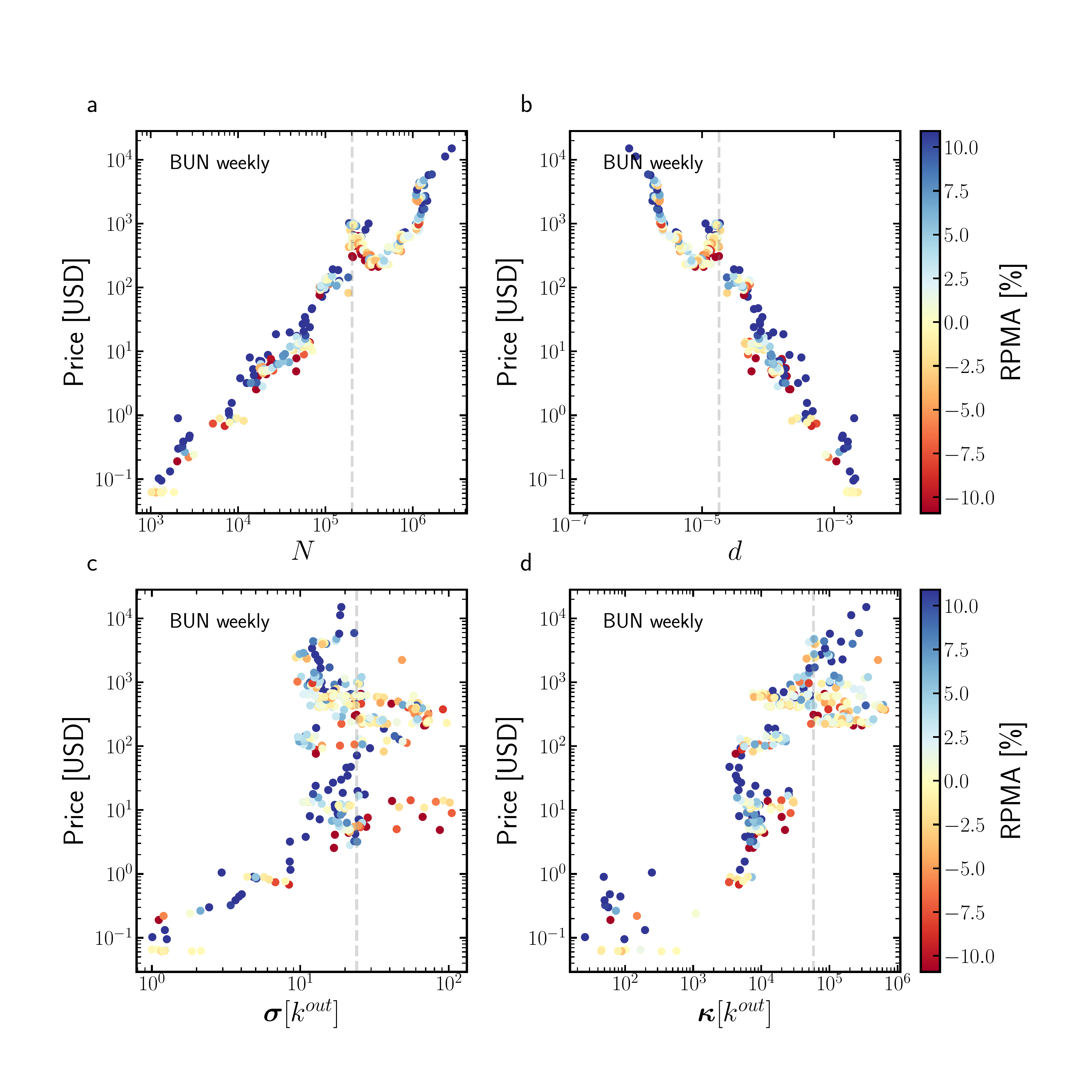}
\caption{Correlation between the Bitcoin price in USD, the basic statistics (number of nodes and link density - top panels) and the moments of the out-degrees distribution (bottom panels) for the BUN at the weekly time scale. Additionally, each dot representing an observation is coloured according to the value of the Ratio between the current Price and its Moving Average (RPMA) indicator. The vertical, dashed line coincides with the bankruptcy of Mt. Gox. Purely structural quantities are correlated with exogenous quantities as the Bitcoin price; see, for example, the evolution of the out-degrees standard deviation whose larger values (observable \emph{after} the Mt. Gox failure) correspond to price drops.}
\label{fig:2}
\end{figure*}

\paragraph*{Degree distributions.} Generally speaking, both out- and in-degrees are characterised by heavy-tailed distributions, indicating that a large number of low-connected nodes co-exists with few hubs whose degree is several order of magnitudes larger. A visual inspection of the functional form of the degrees distributions suggests the latter ones to follow a power-law \cite{bovet2019evolving,Baumann2014ExploringTB}; to test this hypothesis the authors in \cite{bovet2019evolving} employed an algorithm based on a double Kolmogorov-Smirnov statistical test \cite{bauke2007parameter}: what emerges is that the hypothesis above cannot be rejected, at a $0.05$ confidence level, for almost half of the considered snapshots.

Of particular interest is the evolution of the out-degrees standard deviation, especially for what concerns its informativeness about exogenous events. As an example, let us consider the failure in February 2014 of Mt. Gox, a quasi-monopolist exchange market at the time. Such an event deeply affected the overall Bitcoin structure: the percentage of snapshots for which the null hypothesis (i.e. the out-degrees distribution follows a power-law) can be rejected amounts at $\simeq50\%$ \emph{before} February 2014 while it drops to $\simeq25\%$ \emph{afterwards}.

The authors in \cite{bovet2019evolving} also argue that the presence of heavy-tailed distributions may be explained by a mechanism similar to the preferential attachment one: new, or occasional, users `preferentially' connect to already well-connected nodes (exchange markets, utility providers, etc.), thus leading to the formation of super-hubs. Elsewhere it has been argued that the related mechanism known as `fittest-gets-richer' or `good-gets-righer' \cite{bianconi2001bose} may be also at work, the computational resources of a node playing the role of its fitness \cite{javarone2018from}.\\

\paragraph*{Bitcoin structure versus Bitcoin price.} The result concerning the evolution of the out-degrees distribution suggests that the Bitcoin network structure indeed brings the signature of exogenous events. As, in this case, the non-structural quantity \emph{par excellence} is represented by the \emph{price} of our currency, it may be of interest to inspect the presence of correlations between the evolution of the former one and the evolution of purely topological quantities: the justification for such an analysis rests upon the simple consideration that the price of a cryptocurrency is ultimately related to the behaviour of users whose `network' activity translates into that of establishing connections with other nodes - whence our expectation to find some traces of the aforementioned correlations.

The simplest analysis to carry out is that of scattering the network size and the network link density versus the Bitcoin price (in USD). As shown in fig. \ref{fig:2}, a clear trend appears, indicating that the price and the size $N$ (the link density $d$) are overall positively (negatively) correlated throughout the entire Bitcoin history. Notice, however, the trend inversion that can be appreciated immediately after the Mt. Gox failure: it is a consequence of the prolonged price decrease observed in 2014-2016, during which the network size has increased of (almost) one order of magnitude.

To further confirm the presence of a double regime, the authors in \cite{bovet2019evolving} have inspected the correlation between the moments of out-degrees distribution and the Bitcoin price over time. To this aim, the so-called Price and its Moving Average (RPMA) indicator has been adopted, defined as

\begin{equation}
\text{RPMA}_t=100\log_{10}\left(\frac{P_t}{\frac{1}{\tau}\sum^{t-1}_{s=t-1-\tau} P_s}\right)
\end{equation}
with $\tau$ representing a tunable temporal parameter. As shown in \cite{bovet2019evolving}, the standard deviation and the kurtosis diverge as the network size grows larger than the value observed in correspondence of the Mt. Gox failure, thus confirming the `two regimes hypothesis'. Moreover, as fig. \ref{fig:2} shows, larger values of the aforementioned moments (observed \emph{after} the Mt. Gox failure) correspond to price drops, while temporal snapshots corresponding to smaller values of the same quantities seem to be characterised by price increases.

A multivariate Granger test \cite{granger1969investigating} has been also carried out to unveil possible lagged correlations hidden in the data (see fig. 5 in \cite{bovet2019evolving}). To this aim, data have been split in two sub-samples, i.e. 2010-2013 and 2014-2017, and the number of nodes $N$, the number of links $L$ and the higher moments of the empirical (out- and in-) degrees distributions have been put in relation with the log-returns of the Bitcoin price (in USD), within each sub-sample. To sum up, when the BUN is considered at the weekly time scale, a positive feedback loop occurs between $N$ and the price log-returns, whereas at the daily time scale a price increase predicts an increase of the number of nodes $N$ but the viceversa is no longer true. The causality structure is consistent within the two sub-samples.\\

\begin{figure*}[t!]
\centering
\includegraphics[width=\textwidth]{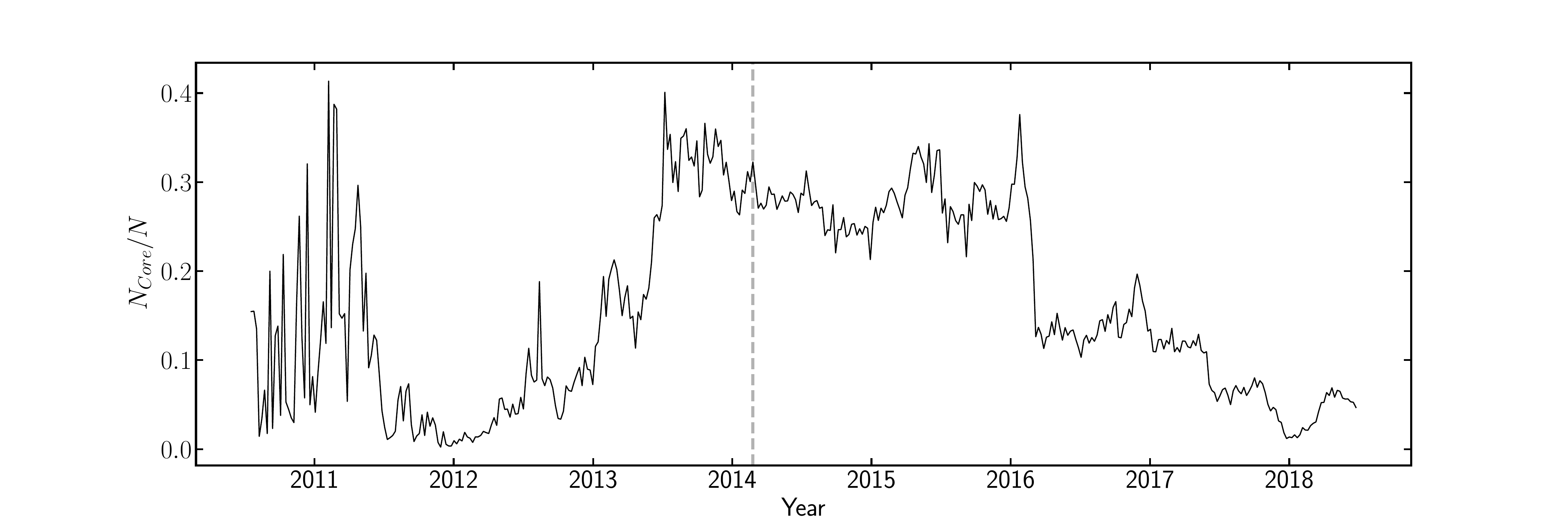}
\caption{Evolution of the percentage of nodes belonging to the core portion of the BUN at the weekly time scale. During the biennium 2012-2013 the core portion of the BUN steadily rises until it reaches $\simeq30\%$ of the network; afterwards, during the biennium 2014-2015, it remains quite constant; then, during the last two years covered by our data set (i.e. 2016-2018), the core portion of the BUN shrinks and the percentage of nodes belonging to it goes back to the pre-2012 values. The vertical, dashed line coincides with the bankruptcy of Mt. Gox.}
\label{fig:3}
\end{figure*}

\paragraph*{Analysis of the BUN mesoscale structure.} Let us now revise the results concerning the mesoscale structure of the BUN. A recently proposed method \cite{de2019detecting} based on the \emph{surprise} score function was adopted by the authors of \cite{vallarano2020exploring} to assess the statistical significance of a peculiar mesoscale organization, known as \emph{core-periphery} structure. According to the interpretation proposed in \cite{de2019detecting}, revealing the core-periphery structure by minimising the surprise means individuating the partition that is least likely to be explained by the null model known as \emph{Random Graph Model} (RGM) with respect to the null model known as \emph{Stochastic Block Model} (SBM) - see also Appendix A. As fig. \ref{fig:3} shows, a core-periphery structure is indeed present: more precisely, during the biennium 2014-2015 the core size amounts at $\simeq30\%$ of the total network size; after 2016, instead, it seems to shrink back to 2010-2013 values. The presence of a core-periphery structure indicates that the BUN is characterised by subgraphs with very different link densities - an evidence that cannot be accounted for by a model defined by just one global parameter, as the one characterising the RGM.

A deeper inspection of the BUN core-periphery structure reveals it to be even richer: in fact, the core portion of the BUN is, actually, the strongly-connected component (SCC) of a \emph{bow-tie} structure whose remaining portions (e.g. the IN and OUT components) compose the BUN periphery \cite{vallarano2020exploring}. More specifically, while the SCC is the set of nodes that are mutually reachable (i.e. a directed path from any node to any other node, within the SCC, exists), the IN and OUT components are respectively defined as the set of nodes from which the SCC can be reached and the set of nodes that can be reached from the SCC. Hence, the picture provided by the evolution of the core-periphery structure can be further refined as follows: since 2016 both the SCC and the OUT-component shrink while the IN-component becomes the dominant portion of the network \cite{vallarano2020exploring}. Other SCCs are visible but their size is negligibly small with respect to the largest one, a finding seemingly indicating that they are, in fact, single nodes pointing to (or pointed by) hubs.

\begin{figure*}[t!]
\centering
\includegraphics[width=\textwidth]{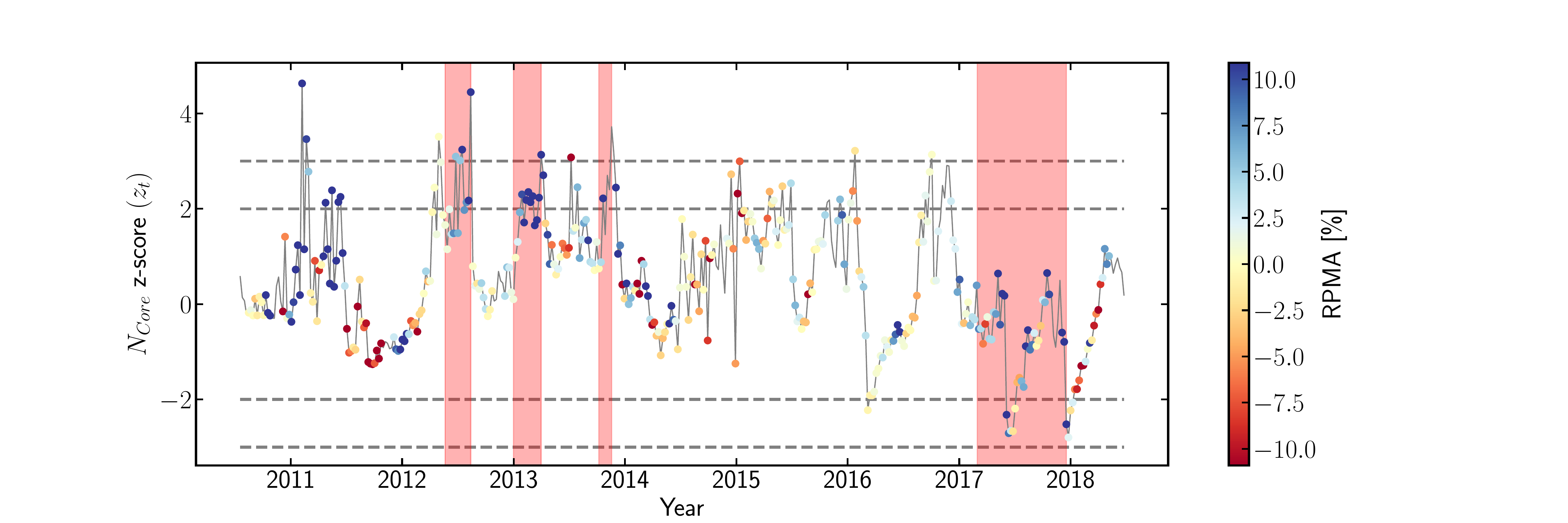}\\
\includegraphics[width=\textwidth]{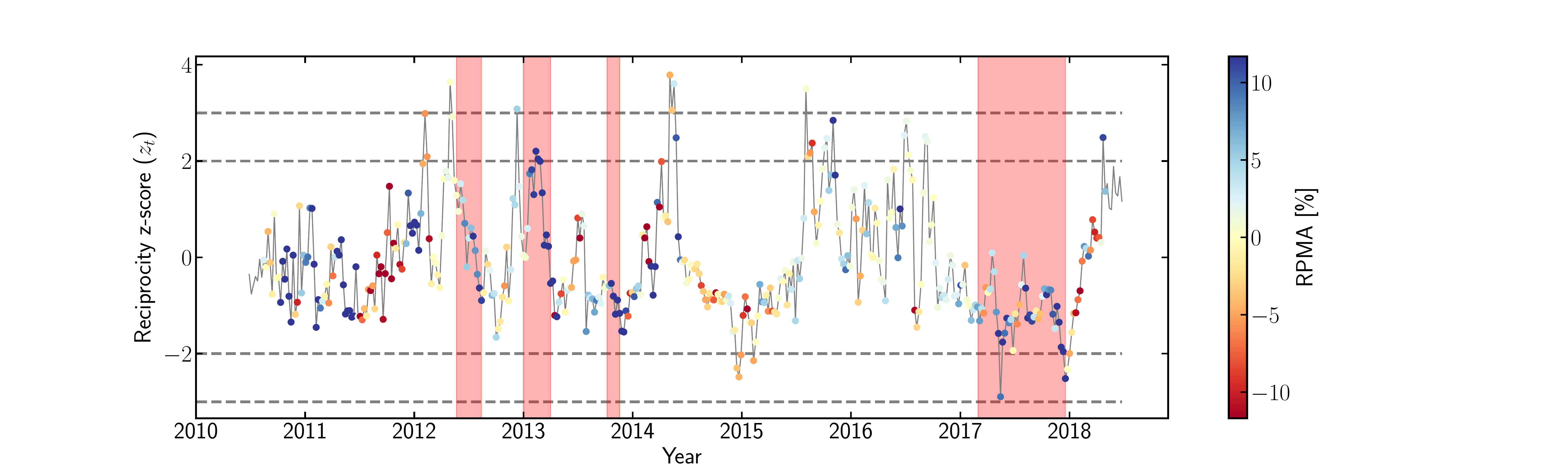}
\caption{Evolution of the temporal $z$-score for the number of core nodes (top panel) and for the reciprocity (bottom panel), for the BUN weekly representation. Shaded areas individuate the so-called \emph{bubbles}, i.e. periods of price increase according to \cite{wheatley2018bitcoin}. Additionally, each dot representing an observation is coloured according to the value of the Ratio between the current Price and its Moving Average (RPMA) indicator. Overall, the two trends show some similarities as peaks are clearly visible in correspondence of the so-called \emph{bubbles}, identified by the shaded areas (see also \cite{wheatley2018bitcoin}). Interestingly, these values are significant in a statistical sense, as the temporal z-scores reach values $z^{(t)}\geq2$.}
\label{fig:4}
\end{figure*}

An additional analysis, aimed at better quantifying the extent to which a generic, purely topological quantity $X$ and the Bitcoin price are related, can be carried out by plotting the evolution of the temporal z-score

\begin{equation}
z_X^{(t)}=\frac{X^{(t)}-\overline{X}}{s_X}    
\end{equation}
where $\overline{X}=\sum_t\frac{X^{t}}{T}$ is the mean over a sample of values covering the period $T$ before time $t$ - in our case, the year before $t$ - and $s_X=\sqrt{\overline{X^2}-\overline{X}^2}$ is the corresponding standard deviation. For example, the choice $X=\sigma[k^{out}]$ allows price drawdowns to be revealed and, in some cases, anticipated \cite{bovet2019evolving}: in the triennium 2010-2012, and after 2017, the price grows as $z_{\sigma[k^{out}]}^{(t)}$ increases while drawdowns appear in periods during which $z_{\sigma[k^{out}]}^{(t)}$ decreases. Other possible choices are $X=N_{core}$ and $X=r$, i.e. the number of core nodes and the network reciprocity, defined as $r=\frac{\sum_i\sum_{j(\neq i)}a_{ij}a_{ji}}{\sum_i\sum_{j(\neq i)}a_{ij}}$, i.e. as the percentage of links having a `partner' pointing in the opposite direction. The evolution of the temporal z-score for the two aforementioned quantities is shown in fig. \ref{fig:4}. Overall, the two trends show some similarities, being characterised by peaks in correspondence of the so-called \emph{bubbles}, i.e. periods of `unsustainable' price growth \cite{wheatley2018bitcoin}: interestingly, such periods are characterised by values of the inspected topological quantities which are significant also in a statistical sense, as the value of the corresponding temporal z-score proves (in fact, $z^{(t)}\geq2$ in both cases). Moreover, peaks are also revealed in the triennium 2014-2016, thus signalling some kind of `activity' missed by purely financial indicators (e.g. the RPMA).

\subsection*{The Bitcoin Lightning Network}

Let us now move to review the results concerning the BLN. In what follows we will focus on the daily-block snapshot representation.\\

\begin{figure*}[t!]
\centering
\includegraphics[width=\textwidth]{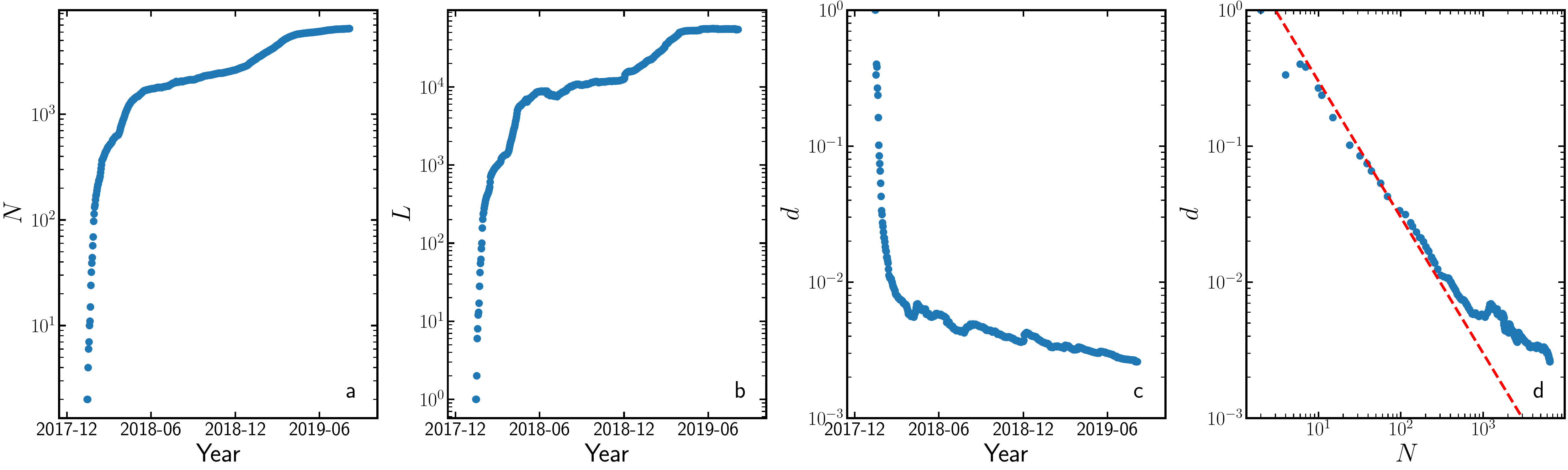}
\caption{Evolution of the total number of nodes $N$, total number of links $L$ and link density $d=\frac{2L}{N(N-1)}$ for the BLN daily-block snapshot representation. As for the BAN and the BUN, the position $d\sim N^{-1}$ well describes the link density dependence on $N$, at least for the snapshots for which $N\leq10^3$. Adapted from \cite{lin2020lightning}.}
\label{fig:5}
\end{figure*}

\paragraph*{Basic statistics.} As observed for the BAN and the BUN, both the number of nodes $N$ and the number of links $L=\sum_i\sum_{j(>i)}b_{ij}$ of the BLN increase steadily in time while it becomes sparser. Interestingly, however, the evolution of the BLN link density seems to point out the presence of two regimes: as fig. \ref{fig:5} shows, during the first phase (i.e. $N\leq10^3$) $L$ increases linearly in $N$ and the link density is well described by the functional dependence $d\sim N^{-1}$; afterwards, the link density decrease slows down, seemingly indicating that $L$ has started to grow in a super-linear fashion with respect to $N$.\\

\paragraph*{Analysis of the BLN mesoscale structure.} Although blockchain-based systems are designed to get rid of the presence of a central authority that checks the validity of the exchanges between nodes - transactions, in the case of cryptocurrencies - and authorizes them, the authors in \cite{lin2020lightning} have shown that centralization may still be recovered at a purely structural level. More precisely, the authors in \cite{lin2020lightning} have considered two different sets of quantities. First, they have computed the Gini coefficient

\begin{equation}
G_c=\frac{\sum_{i=1}^N\sum_{j=1}^N|c_i-c_j|}{2N\sum_{i=1}^N{c_i}}
\end{equation}
for four centrality measures, i.e. the \emph{degree}, \emph{closeness}, \emph{betweenness} and \emph{eigenvector centrality} (respectively indicated with the symbols $c_i=k^c_i,c^c_i,b^c_i,e^c_i$ - see also Appendix B and \cite{wang2018improved}) and plotted it versus the number of nodes. As shown in \cite{lin2020lightning} $G_c$ increases for three measures out of four, i.e. the degree, betweenness and eigenvector dentrality but not for the closeness centrality whose trend remains basically flat. Since the Gini coefficient quantifies the (un)evenness of a distribution, this result points out that the centrality of nodes is more and more unevenly distributed. A concrete example is provided by the value $G_{k^c}$ reaching $\simeq 0.8$ in the last snapshot of our data set: this value is compatible with the picture of a network where the $90\%$ of connections are incident to the 10\% of nodes. In other words, nodes exist playing the role of \emph{hubs}, i.e. vertices with a large number of connections, that are crossed by a large percentage of paths and that are connected to other well-connected nodes.

Additionally, the authors in \cite{lin2020lightning} have also computed the so called \textit{centralization indices}, encoding the comparison between the structure of a given network and that of a reference network, i.e. the `most centralized' structure - see also Appendix B. For what concerns the degree, closeness and betweenness centrality, it is the \emph{star graph}: for what concerns the eigenvector index, the star graph does not represent the maximally-centralized structure, although it is kept for the sake of homogeneity with the other quantities. The evolution of the centralization indices indicates that the BLN is not evolving towards a star graph (indeed, a too simplistic picture to faithfully describe the BLN topology) but towards a suitable generalization of it, i.e. the core-periphery structure \cite{lin2020lightning} (see also later). Incidentally, the presence of a core-periphery structure is compatible with the aforementioned even distribution of the closeness centrality since, by definition, the closeness of a core node does not differ much from the closeness of a periphery node.

\begin{figure*}[t!]
\centering
\includegraphics[width=\textwidth]{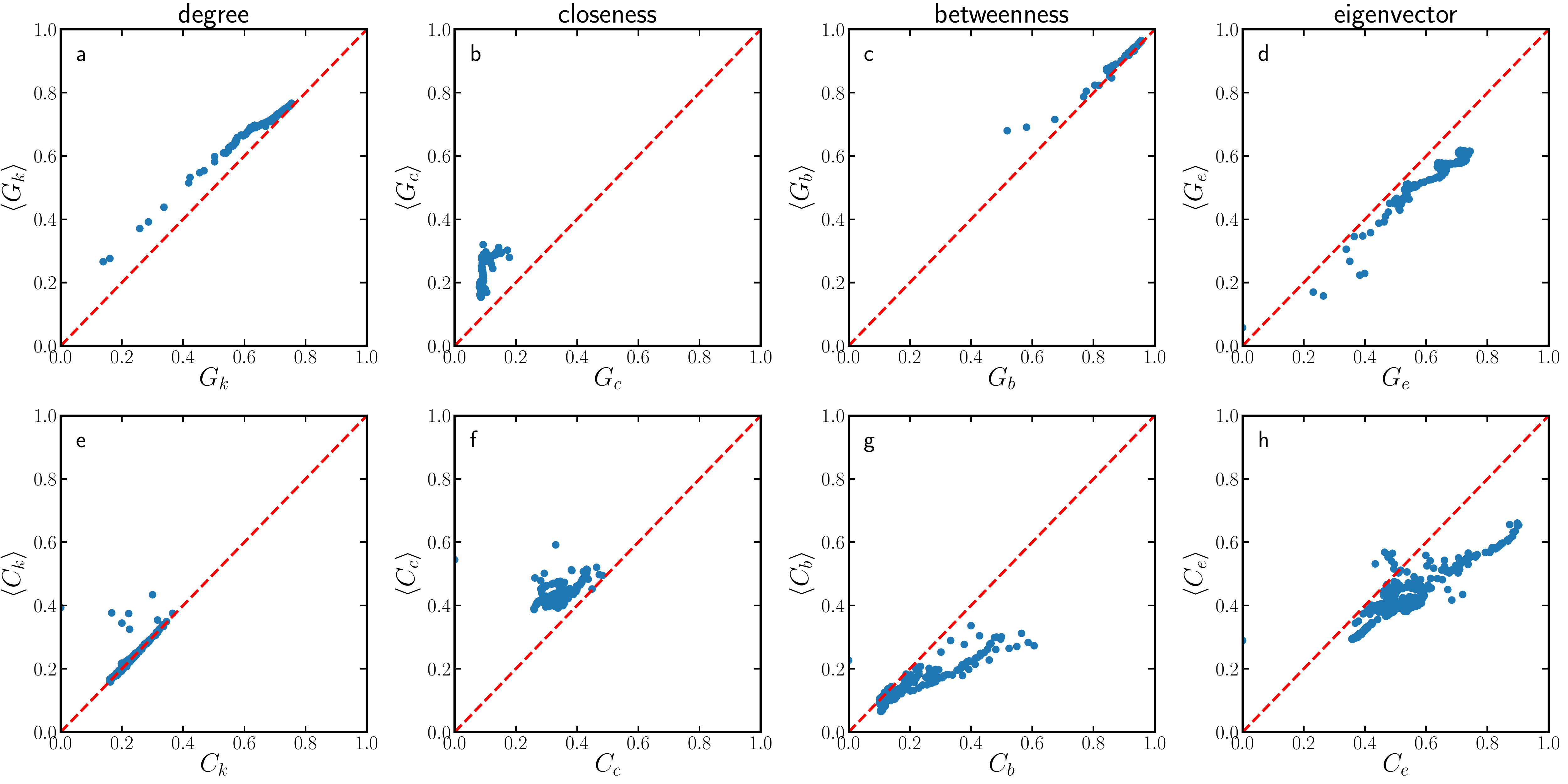}
\caption{Top panels: comparison between the observed Gini index for the degree, closeness, betweenness and eigenvector centrality (x-axis) and their expected value, computed under the UBCM (y-axis) for the BLN daily-block snapshot representation. Bottom panels: comparison between the observed degree-, closeness-, betweenness-and eigenvector-centralization measures and their expected value computed under the UBCM. Once the information contained into the degree sequence is properly accounted for, a (residual) tendency to centralization is still visible, letting the picture of a network characterised by some kind of more-than-expected `star-likeness' emerge.}
\label{fig:6}
\end{figure*}

The authors in \cite{lin2020lightning} have also benchmarked the observations concerning the evolution of the centrality and the centralization indices with the predictions, for the same quantities, output by the maximum-entropy null model known as \emph{Undirected Binary Configuration Model} (UBCM - see also Appendix C). To this aim, they have explicitly sampled the ensembles of networks induced by the UBCM \cite{park2004statistical,squartini2011analytical} and compared the ensemble average of the quantities of interest with the corresponding empirical values. From a merely technical point of view, the authors adopt an iterative, reduced algorithm to solve the system of equations defining the UBCM, i.e.

\begin{equation}
k_i(\mathbf{A})=\sum_{j(\neq i)=1}^N\frac{x_ix_j}{1+x_ix_j},\:\forall\:i\quad\Longrightarrow\quad x^{(n)}_k=\frac{k(\mathbf{A})}{\sum_{k'}f(k')\left[\frac{x^{(n-1)}_{k'}}{1+x^{(n-1)}_{k}x^{(n-1)}_{k'}}\right]-\frac{x^{(n-1)}_{k}}{1+(x^2_{k})^{(n-1)}}},\:\forall\:k
\end{equation}
a choice allowing them to solve it within tens of seconds even for configurations with millions of nodes \cite{vallarano2020exploring} - see also Appendix C. As fig. \ref{fig:6} shows, such a comparison reveals that the UBCM tends to overestimate the values of the Gini index for the degree, the closeness and the betweenness centrality and to underestimate its values for the eigenvector centrality. This seems to point out a non-trivial (i.e. not reproducible by just enforcing the degrees) tendency of well-connected nodes to establish connections among themselves - likely, with nodes having a smaller degree attached to them: such a disassortative structure could explain the less-than-expected level of unevenness characterising the other centrality measures, as each of the nodes behaving as the `leaves' of the hubs would basically have the same values of degree, closeness and betweenness centrality.

\begin{figure*}[t!]
\centering
\includegraphics[width=\textwidth]{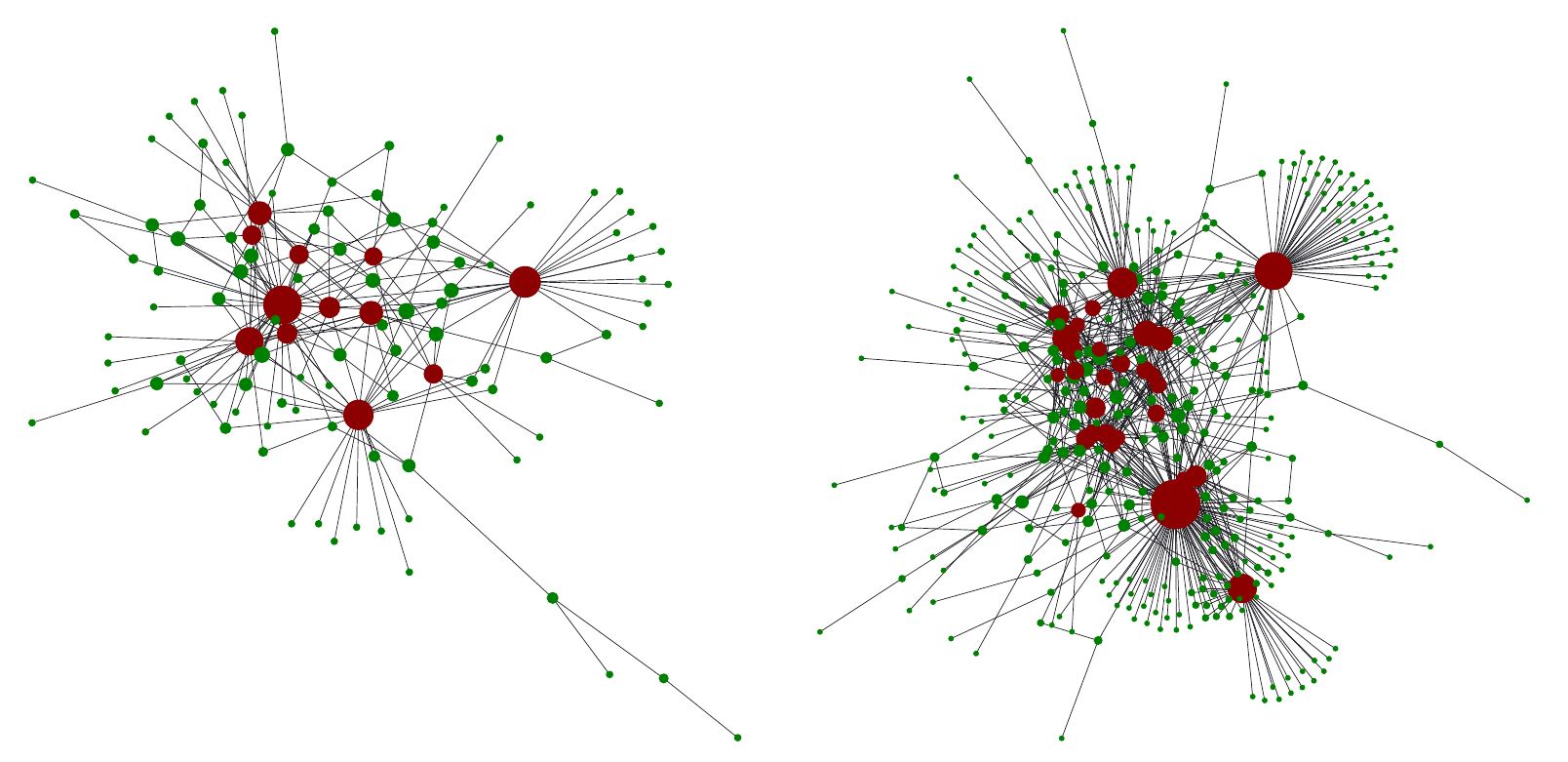}
\caption{Core-periphery structure of the BLN daily-block representation on day 17 (left panel) and on day 35 (right panel), with core-nodes drawn in red and periphery-nodes drawn in green. Adapted from \cite{lin2020lightning}.}
\label{fig:7}
\end{figure*}

For what concerns the analysis of the centralization indices, fig. \ref{fig:6} shows that the UBCM underestimates both the betweenness- and the eigenvector-centralization indices: in other words, a tendency to centralization `survives' even after the information encoded into the degrees is properly accounted for, letting the picture of a network characterised by some kind of more-than-expected `star-likeness' emerge. This observation can be better formalised by analysing the BLN mesoscale structure via the optimization of surprise: as observed for the BUN, a core-periphery structural organization, whose statistical significance increases over time, indeed emerges \cite{lin2020lightning} (see also fig. \ref{fig:7}).

In \cite{vallarano2020exploring}, the authors have also adapted the iterative, reduced algorithm cited above for the resolution of the \emph{Directed Binary Configuration Model} (DBCM - see also Appendix C).\\

\paragraph*{A quick look at the weighted structure of the BLN.} Having a quick look at the weighted structure of the BLN leads to two notable observations: both the \emph{total amount of exchanged bitcoins} and the \emph{unevenness of their distribution} increase. This trend is confirmed by the evolution the Gini coefficient whose value reaches $0.9$ for the last snapshots of our data set. On average, across the entire period, about the $10\%$ ($50\%$) of nodes holds the $80\%$ ($99\%$) of the bitcoins at stake in the network \cite{lin2020lightning}.

\section*{Discussion}

The public availability of the complete history of Bitcoin transactions allows researchers to analyse the \emph{structure} characterising different transaction networks, to inspect the inter-dependency between its dynamics and that of the Bitcoin \emph{price} and to gain insight into the \emph{behaviour} of Bitcoin users; still, the understanding of the mechanisms driving the joint evolution of the three entities above remains far from being complete.

This paper provides an overview of the most recent results on the topic achieved in the last years. One of the main messages concerns the possibility to retrieve signals of exogenous events by analysing the blockchain-induced transaction networks: the best example is provided by the failure of Mt. Gox in 2014, an event deeply affecting both the Bitcoin Address Network and the Bitcoin User Network structure. From this point of view, out-degrees have been found to represent particularly informative properties: higher moments of the out-degrees distribution (as the standard deviation, the skewness and the kurtosis) diverge as the network size grows larger than the value observed in correspondence of the Mt. Gox failure; besides, the out-degrees heterogeneity rises during periods of price decline (and vice-versa).

Such a result is further refined by a Granger causality analysis, revealing that during the triennium 2010-2012 an increase of the out-degrees standard deviation \emph{causes} a price decline \cite{bovet2019evolving}. This finding, in turn, suggests a sort of behavioural explanation for the price dynamics displayed during the early stages of Bitcoin: during periods in which the price continuously increases, an increasing number of traders is, in turn, attracted to the system; the latter ones, likely performing only few transactions, link to the network hubs (usually exchange markets) that gain a large number of connections over the course of the weeks, thus causing the price to rise even more.

Interestingly, the analysis of the Bitcoin Lightning Network reveals the same trends observed for the BAN and the BUN, as the emergence of an uneven distribution of the centrality and the wealth of nodes and of a statistically-significant core-periphery structure. These results suggest the tendency of the Bitcoin `Layer 2' network to become less distributed, a process having the undesirable consequence of making this off-chain payment network less resilient to random failures, malicious attacks, etc. The emergence of hubs may be a consequence of the way the BLN is designed: as a route through the network must be found and longer routes are more expensive (fees are present for the gateway service provided by intermediate nodes), any two BLN users will search for a short(est) path; at the same time, nodes have the incentive to become as central as possible, in order to maximize the transaction fees they may earn. Hubs may, thus, have emerged as a consequence of the collective action of users following one of the two aforemenioned behaviours - not surprisingly, since the very beginning of the BLN history. For what concerns hubs interconnectedness, previous results have shown that mechanisms aiming at maximizing the centrality of agents yield a core-periphery structure (regardless of the notion of centrality the agents attempt to maximize) \cite{konig2014nestedness,konig2010from}. As a last observation, we also notice that the presence of `centrality hubs' seems to be at the origin of another peculiarity of the BLN structural, i.e. its small-world -ness (a feature already revealed by previous studies \cite{rohrer2019discharged}).

The results reviewed in this article ultimately - and consistently - point out a tendency to centralization that has been observed in the Bitcoin structure at different levels \cite{lin2020lightning,gervais2014bitcoin}, an evidence that deserves to be investigated in greater detail. A natural extensions of the present work moves in the direction of analysing the \emph{weighted} counterparts of the three constructs considered here. Of particular interest would be the analysis of the weighted centrality and centralization indices considered in \cite{lin2020lightning}, whose outcome would help clarifying to what extent the evidence that binary and weighted quantities are usually correlated in financial systems holds true for cryptocurrencies as well. Other promising avenues of research concern the analysis of different cryptocurrencies as well as other blockchain-based systems, to understand if the same mechanisms shaping the Bitcoin structure are at work also elsewhere.

\section*{Acknowledgements}

C.J.T. acknowledges financial support by the University of Zurich through the University Research Priority Program on Social Networks. T.S. acknowledge support from the EU project SoBigData-PlusPlus (grant no. 871042). The authors acknowledge A. Bovet, C. Campajola, F. Mottes, V. Restocchi, J.-H. Lin for useful discussions. 

\section*{Author Contributions}

All authors wrote, reviewed and approved the manuscript.

\section*{Competing Interests}

The authors declare no competing financial interests.

\section*{Data Availability}

Data concerning the entire Bitcoin transaction history is publicly available at the address \emph{https://www.blockchain.com/}. By synchronising the desktop client Bitcoin Core, available at \emph{https://bitcoin.org/en/download}, anybody can have a local copy of the entire transaction history.

\section*{Appendix A - Detecting core-periphery structures by surprise}

The `generalised' star graph structure also known as \emph{core-periphery structure} is defined by a densely-connected core of nodes surrounded by a periphery of loosely-connected vertices. In order to check for its presence, we implement a recently-proposed approach \cite{de2019detecting}, prescribing to minimise the score function known as \emph{bimodular surprise} and reading

\begin{equation}
\mathscr{S}_\parallel=\sum_{i\geq l_\bullet^*}\sum_{j\geq l_\circ^*}\frac{\binom{V_\bullet}{i}\binom{V_\circ}{j}\binom{V-(V_\bullet+V_\circ)}{L-(i+j)}}{\binom{V}{L}}
\end{equation}
which is nothing else that the multinomial version of the \emph{surprise}, originally proposed to carry out a \emph{community detection} exercise \cite{nicolini2016modular}. The presence of three different binomial coefficients allows three different `species' of links to be accounted for: the binomial coefficient $\binom{V_\bullet}{i}$ enumerates the number of ways $i$ links can redistributed \emph{within} the first module (e.g. the core portion), the binomial coefficient $\binom{V_\circ}{j}$ enumerates the number of ways $j$ links can redistributed \emph{within} the second module (e.g. the periphery portion) and the binomial coefficient $\binom{V-(V_\bullet+V_\circ)}{L-(i+j)}$ enumerates the number of ways the remaining $L-(i+j)$ links can be redistributed \emph{between} the first and the second module, i.e. over the remaining $V-(V_\bullet+V_\circ)$ node pairs; the values $i$ and $j$ are bounded by the values $V_\bullet$ and $V_\circ$, although the sum $i+j$ can range between $l_\bullet^*+l_\circ^*$ and the minimum between $L$ and $V_\bullet+V_\circ$.

From a technical point of view, $\mathscr{S}_\parallel$ is the p-value of a multivariate hypergeometric distribution, describing the probability of $i+j$ successes in $L$ draws (without replacement), from a finite population of size $V$ that contains exactly $V_\bullet$ objects with a first specific feature and $V_\circ$ objects with a second specific feature.

\section*{Appendix B - Centrality and centralization measures}

Indices measuring the centrality of a node aim at quantifying the `importance' of a node in a network, according to some specific topological property. Among the measures proposed so far, of particular relevance are the \emph{degree centrality}, the \emph{closeness centrality}, the \emph{betweenness centrality} and the \emph{eigenvector centrality} \cite{wang2018improved}:

\begin{itemize}
\item the \emph{degree centrality} of node $i$ coincides with the degree of node $i$, i.e. the number of its neighbours, normalised by the maximum attainable value, i.e. $N-1$:

\begin{equation}\label{equation1}
k^c_{i}=\frac{k_i}{N-1}.
\end{equation}

From the definition above, it follows that the most central node, according to the degree variant, is the one connected to all the other nodes;

\item the \emph{closeness centrality} of node $i$ is defined as

\begin{equation}\label{equation3}
c^c_{i}=\frac{N-1}{\sum_{j(\neq i)=1}^Nd_{ij}}
\end{equation}
where $d_{ij}$ is the topological distance between nodes $i$ and $j$, i.e. the length of the shortest path(s) connecting them. From the definition above, it follows that the most central node, according to the closeness variant, is the one lying at distance 1 from each other node;

\item the \emph{betweenness centrality} of node $i$ is given by

\begin{equation}\label{equation2}
b^c_{i}=\sum_{s(\neq i)=1}^N\sum_{t(\neq i,s)=1}^N\frac{\sigma_{st}(i)}{\sigma_{st}}
\end{equation}
where $\sigma_{st}$ is the total number of shortest paths between node $s$ and $t$ and $\sigma_{st}(i)$ is the number of shortest paths between nodes $s$ and $t$ that pass through node $i$. From the definition above, it follows that the most central node, according to the betweenness variant, is the one lying `between' any two other nodes;

\item the \emph{eigenvector centrality} of node $i$, $e^c_{i}$, is defined as the $i$-th element of the eigenvector corresponding to the largest eigenvalue of the binary adjacency matrix (whose existence is ensured by the Perron-Frobenius theorem). According to the definition above, a node with large eigenvector centrality is connected to other `well connected' nodes \cite{Baumann2014ExploringTB}.
\end{itemize}

The centrality indices defined above provide a rank of the nodes of a network. Sometimes, however, it is useful to compactly describe a certain network structure in its entirety. To this aim, a family of indices, known as \emph{centralization indices}, has been defined. In mathematical terms, any centralization index reads

\begin{equation}
C_c=\frac{\sum_{i=1}^N(c^*-c_i)}{\max\{\sum_{i=1}^N(c^*-c_i)\}}
\end{equation}
where $c^*=\max\{c_i\}_{i=1}^N$ represents the empirical, maximum value of the chosen centrality measure (i.e. computed on the network under consideration) and the denominator is calculated over a benchmark graph, defined as the one providing the maximum attainable value of the quantity $\sum_{i=1}^N(c^*-c_i)$. The most centralized structure, according to the degree, closeness and betweenness centrality is the \emph{star graph}, in correspondence of which one finds that

\begin{itemize}
\item $\sum_{i=1}^N(k^*-k^c_i)=(N-1)(N-2)$;
\item $\sum_{i=1}^N(c^*-c^c_i)=\frac{(N-1)(N-2)}{2N-3}$;
\item $\sum_{i=1}^N(b^*-b^c_i)=\frac{(N-1)^2(N-2)}{2}$.
\end{itemize}

For what concerns the \emph{eigenvector-centralization} index, the star graph does not represent the maximally centralized structure; however, for the sake of comparison with the quantities above, the authors in \cite{lin2020lightning} have calculated it on a star graph as well, in correspondence of which one finds that $\sum_{i=1}^N(e^*-e^c_i)=(\sqrt{N-1}-1)(N-1)/(\sqrt{N-1}+N-1)$.

\section*{Appendix C - An iterative method to solve null models}

The significance of any result can be assessed only after a comparison with a properly-defined benchmark (or null) model. To this aim, one can consider the \emph{Exponential Random Graph} (ERG) framework. Generally speaking, the problem to be solved in order to define a benchmark model within such a framework reads

\begin{equation}
\max_{P}\left\{S[P]-\sum_{i=0}^M\theta_i\left[\sum_{\mathbf{A}}P(\mathbf{A})C(\mathbf{A})-\langle C_i\rangle\right]\right\}
\end{equation}
where 

\begin{equation}
S[P]=-\sum_\mathbf{A}P(\mathbf{A})\ln P(\mathbf{A})
\end{equation}
is \emph{Shannon entropy} and $\vec{C}(\mathbf{A})$ is an $M$-dimensional vector of constraints representing the information defining the benchmark - notice that $C_0=\langle C_0\rangle=1$ sums up the normalization condition of the probability distribution $P(\mathbf{A})$. The solution to the problem above reads

\begin{equation}
P(\mathbf{A},\vec{\theta})=\frac{e^{-H(\mathbf{A},\vec{\theta})}}{Z(\vec{\theta})}
\end{equation}
with $Z(\vec{\theta})=\sum_\mathbf{A}P(\mathbf{A},\vec{\theta})$ representing the \emph{partition function} and $H(\mathbf{A},\vec{\theta})=\vec{\theta}\cdot\vec{C}(\mathbf{A})$ representing the \emph{Hamiltonian}, i.e. the functions summing up the normalization condition and the imposed constraints, respectively.\\

\paragraph*{The Undirected Binary Configuration Model (UBCM).} In case the Undirected Binary Configuration Model (UBCM) is chosen as a benchmark, the Hamiltonian reads 

\begin{equation}
H(\mathbf{A},\theta)\equiv\vec{\theta}\cdot\vec{k}(\mathbf{A})=\sum_{i=1}^N\sum_{j(>i)=1}^N(\theta_i+\theta_j)a_{ij}
\end{equation}
a position leading to the probability function $P(\mathbf{A})=\prod_i\prod_{j(>i)}p_{ij}^{a_{ij}}(1-p_{ij})^{1-a_{ij}}$ with $p_{ij}^\text{UBCM}\equiv\frac{e^{-(\theta_i+\theta_j)}}{1+e^{-(\theta_i+\theta_j)}}\equiv\frac{x_ix_j}{1+x_ix_j}$. The unknown parameters can be estimated by invoking a second maximization principle, i.e. the maximization of the likelihood function. The latter is defined as 

\begin{equation}
\mathcal{L}(\vec{x})=\ln P(\mathbf{A}|\vec{x})
\end{equation}
and needs to be optimised with respect to the vector of unknown parameters $\vec{x}$. Remarkably, whenever the probability distribution is exponential (as the one deriving from Shannon entropy maximization), the likelihood maximization leads to the system $\langle\vec{C}\rangle=\vec{C}(\mathbf{A})$ to be solved, that in the UBCM case reads

\begin{equation}
k_i(\mathbf{A})=\sum_{j(\neq i)=1}^N\frac{x_ix_j}{1+x_ix_j},\:\forall\:i.
\end{equation}

In order to solve the system above, the iterative recipe

\begin{equation}
x_i^{(n)}=\frac{k_i(\mathbf{A})}{\sum_{j(\neq i)=1}^N\left[\frac{x_j^{(n-1)}}{1+x_i^{(n-1)}x_j^{(n-1)}}\right]},\:\forall\:i
\end{equation}
can be employed; naturally, such a recipe needs to be initialised: the values $x_i^{(0)}=\frac{k_i(\mathbf{A})}{\sqrt{2L}},\:\forall\:i$ can be chosen, i.e. the solution to the system of equations defining the UBCM in the sparse case. Let us notice that the computation of the system above can be further sped up, by assigning to the nodes with the same degree $k$ the same value of the hidden variable $x$, i.e.

\begin{equation}
x^{(n)}_k=\frac{k(\mathbf{A})}{\sum_{k'}f(k')\left[\frac{x^{(n-1)}_{k'}}{1+x^{(n-1)}_{k}x^{(n-1)}_{k'}}\right]-\frac{x^{(n-1)}_{k}}{1+(x^2_{k})^{(n-1)}}},\:\forall\:k
\end{equation}
where the sum runs over the \emph{distinct} values of the degrees and $f(k)$ is the number of nodes whose degree is $k$ \cite{garlaschelli2008maximum}.\\

\paragraph*{The Directed Binary Configuration Model (DBCM).} In case the Directed Binary Configuration Model (DBCM) is chosen as a benchmark, the Hamiltonian reads 

\begin{equation}
H(\mathbf{A},\alpha,\beta)\equiv\vec{\alpha}\cdot\vec{k}^{out}(\mathbf{A})+\vec{\beta}\cdot\vec{k}^{in}(\mathbf{A})=\sum_{i=1}^N\sum_{j(\neq i)=1}^N(\alpha_i+\beta_j)a_{ij}
\end{equation}
a position leading to the probability function $P(\mathbf{A})=\prod_i\prod_{j(\neq i)}p_{ij}^{a_{ij}}(1-p_{ij})^{1-a_{ij}}$ with $p_{ij}^\text{DBCM}\equiv\frac{e^{-(\alpha_i+\beta_j)}}{1+e^{-(\alpha_i+\beta_j)}}\equiv\frac{x_iy_j}{1+x_iy_j}$. As for the UBCM, the unknown parameters can be estimated by invoking the maximization of the likelihood function. In the DBCM case, it leads to the system

\begin{eqnarray}
k_i^{out}(\mathbf{A})&=&\sum_{j(\neq i)=1}^N\frac{x_iy_j}{1+x_iy_j},\:\forall\:i\\
\:k_i^{in}(\mathbf{A})&=&\sum_{j(\neq i)=1}^N\frac{x_jy_i}{1+x_jy_i},\:\forall\:i.
\end{eqnarray}

In order to solve the system above, the iterative recipe (originally proposed in \cite{dianati2016maximum} and further refined in \cite{vallarano2020exploring})

\begin{eqnarray}
x_i^{(n)}=\frac{k_i^{out}(\mathbf{A})}{\sum_{j(\neq i)=1}^N\left[\frac{y_j^{(n-1)}}{1+x_i^{(n-1)}y_j^{(n-1)}}\right]},\:\forall\:i\\
\:y_i^{(n)}=\frac{k_i^{in}(\mathbf{A})}{\sum_{j(\neq i)=1}^N\left[\frac{x_j^{(n-1)}}{1+x_j^{(n-1)}y_i^{(n-1)}}\right]},\:\forall\:i
\end{eqnarray}
can be employed. As for the UBCM case, the solutions to the system of equations defining the DBCM in the sparse case, i.e. $x_i^{(0)}=\frac{k_i^{out}(\mathbf{A})}{\sqrt{L}},\:\forall\:i$ and $y_i^{(0)}=\frac{k_i^{in}(\mathbf{A})}{\sqrt{L}},\:\forall\:i$, can be chosen to initialise the recipe above. The computation of the system above can be further sped up, by assigning to the nodes with the same (pair of) out- and in-degrees $(k_i^{out},k_i^{in})$ the same pair of values $(x,y)$ \cite{garlaschelli2008maximum}.

\end{document}